\documentclass[prb,aps,twocolumn,draft,tightenlines,%
showpacs,floatfix,superscriptaddress]{revtex4}
\usepackage{colordvi}
\usepackage{psfig}
\usepackage{amssymb}

\usepackage{amsmath}
\begin{document}
\title{Spin relaxation in semiconductor quantum dots}
\author{J. L. Cheng}
\author{M. W. Wu}%
\thanks{Author to whom correspondence should be addressed}%
\email{mwwu@ustc.edu.cn}
\affiliation{Structure Research Laboratory, University of Science \&%
Technology of China, Academia Sinica,  Hefei, Anhui, 230026, China}
\affiliation{Department of Physics, University of Science \&%
Technology of China, Hefei, Anhui, 230026, China}%
\altaffiliation{Mailing Address.}
\author{C. L\"u}
\affiliation{Department of Physics, University of Science \&%
Technology of China, Hefei, Anhui, 230026, China}
\date{\today}
\begin{abstract}
The spin relaxation time due to the electron-acoustic phonon
scattering  in GaAs quantum dots is studied  after the exact 
diagonalization of the electron Hamiltonian with 
the spin-orbit coupling. Different effects such as the magnetic
field, the quantum dot size and the temperature 
on the spin relaxation time are investigated in detail. Moreover,
we show that the perturbation method widely used in the
literature  is inadequate in accounting for the electron structure
and therefore the spin relaxation time.

\end{abstract}
\pacs{71.70.Ej, 73.21.La,72.25.Rb}

\maketitle
\section{Introduction}
Spin-related phenomena in semiconductors
have attracted much attention recently as they are the key ingredient in
the field of spintronics.\cite{spintronics} 
Among these, the spin-orbit coupling mechanisms in semiconductor quantum
dots (QDs) provide a basis for device applications such as qubits in quantum
computers and have therefore caused much 
interest.\cite{manuel,Gupta,Khaetsii,Hackens,Alexander1,Alexander}
Voskoboynikov {\em et al.} studied  the electron structures  of QDs
by exactly diagonalizing the Hamiltonian with spin-orbit 
coupling.\cite{Voskoboynikov} 
Governale studied the electron structure of few-electron 
interacting QDs  with Rashba spin-orbit coupling by spin 
density functional theory.\cite{Governale}  Val\'in-Rodr\'iguez  {\em et al.} 
investigated spin procession in QDs with the spin-orbit coupling.\cite{manuel}
Besides the effect of the spin-mixing in the electron structure, the
spin-orbit coupling also induces the spin relaxation via further coupling 
with phonons, which alone conserve the spin and therefore are unable
to  cause any spin relaxation. Many 
works calculated the spin relaxation time (SRT)  
due to the spin-orbit coupling induced spin-flip electron-phonon scattering
at zero or very low temperatures.\cite{Woods,Khaetsii,Alexander,Alexander1}
Unlike the electron structure 
calculation,\cite{Voskoboynikov,Governale,manuel} to our
knowledge all works on the SRT are based on perturbation theory where the 
spin-orbit coupling is treated as a perturbation in the Hilbert space
spanned by $H_0$ which does not include the spin-orbit coupling. Moreover
only the lowest few energy levels of $H_0$  are included in the
theory.\cite{Woods,Khaetsii,Alexander,Alexander1} Whether the perturbation
based on the lowest few levels of $H_0$ is adequate remains unchecked.

In the present paper, we investigate the SRT of GaAs
QDs confined in the quantum well by parabolic potentials
by exactly diagonalizing the total Hamiltonian.  
We calculate the SRT due to the scattering with the acoustic phonons
by the Fermi golden rule after 
getting the energy spectra and the wavefunctions from the 
exact diagonalization. 
We find that the perturbation approach is inadequate in
calculating the SRT and therefore it is necessary to reinvestigate the 
accurate SRT via the exact diagonalization approach.
We organize the paper as follows: In Sec.\ II we set up our model and
the Hamiltonian. 
Then in Sec.\ III we present our numerical results: We first compare
the results 
obtained from our exact diagonalization method with those from the perturbation
approach and show that the perturbation method is inadequate in accounting for the 
SRT in Sec.\ III(A). In Sec.\ III(B) we discuss the SRT of a small QD at $T=4$\ K
where only the lowest two energy levels of the total electron Hamiltonian 
after the exact diagonalization account 
for the SRT. Nevertheless at least 12 energy levels of $H_0$  are needed to get these
lowest two energy levels. We then turn to effects of the magnetic
field, the temperature and the 
quantum well width to the SRT in Sec.\ III(C)-(E).
We give our conclusions in Sec.\ IV.

\section{Model and Hamiltonian}
We set up a simplified model to study the spin relaxation in
the QD's which are defined by parabolic potentials $V_c({\bf
  r})=\frac{1}{2}m^{\ast}\omega^2_0{\bf r}^2$  in a quantum well 
of width $a$. A magnetic field ${\bf B}$ is applied along the growth
($z$) direction of the quantum well. The total Hamiltonian is
given by  
\begin{equation}
  \label{eq:total_hamitonian}
  H = H_e+H_{ph}+H_{ep}
\end{equation}
with the electron Hamiltonian  $H_e=H_0+H_{so}$. 
Here $H_0$ is electron Hamiltonian without the spin-orbit coupling:
\begin{equation}
H_0=\frac{{\bf P}^2}{2m^{\ast}}+V_{c}({\bf r}) 
  + H_B \,
\end{equation}
in which ${\bf P} = -i\hbar\mbox{\boldmath$\nabla$} + (e/c){\bf A}$ with
${\bf A} = \frac{B}{2}(-y, x, 0)$ stands for the electron momentum
operator. $m^{\ast}$ is the electron effective 
mass. $H_B=\frac{1}{2}g\mu_BB\sigma_z$ is the
Zeeman energy with \boldmath$\sigma$\unboldmath 
 representing
the Pauli matrices.
$H_{so}=\gamma{\bf h}\cdot\mbox{\boldmath$\sigma$\unboldmath}$ is the
spin-orbit coupling which is the key to the spin flip and spin relaxation.
${\bf h}=[P_x(P_y^2-P_z^2),
 P_y(P_z^2-P_x^2), P_z(P_x^2-P_y^2)]$ is the Dresshauls effective
 magnetic field in the bulk material.\cite{DP} 
In quantum well with small width, $H_{so}$ can  be simplified as 
\begin{equation}
\label{so}
H_{so} = \gamma_c(-P_x\sigma_x + P_y\sigma_y)
\end{equation}
with $\gamma_c=\gamma(\pi/a)^2$.
$H_{ph}$ in Eq.\ (1) is the Hamiltonian for phonons and is
given by $H_{ph}= \sum_{{\bf 
 q}\lambda}\hbar\omega_{{\bf q}\lambda}a^{\dagger}_{{\bf
 q}\lambda}a_{{\bf q}\lambda}$ with $\omega_{{\bf q}\lambda}$ 
standing for the phonon energy spectrum
of branch $\lambda$ and momentum ${\bf q}$.
The electron-phonon scattering is given by
\begin{equation}
\label{ep}
H_{ep} = \sum_{{\bf q}\lambda}M_{{\bf
 q}\lambda}(a^{\dagger}_{{\bf q}\lambda}+a_{{\bf q}\lambda})\exp(i{\bf
q}\cdot{\bf r})
\end{equation}
 with $M_{{\bf q}\lambda}$ being the scattering matrix element.

We diagonalize the electron Hamiltonian $H_e$ in the Hilbert space
$|n,l,\sigma\rangle$
constructed by $H_0= \frac{{\bf P}^2}{2m^{\ast}}+V_{c}({\bf r})+H_B$:
$|\Psi_\ell\rangle=\sum_{nl\sigma}C_{nl\sigma}^\ell |n,l,\sigma\rangle$.
Here $H_0|n,l,\sigma\rangle=E_{n,l,\sigma}|n,l,\sigma\rangle$ with
\begin{equation}
\label{eq:eigen}
\langle {\bf r}|n,l,\sigma\rangle=N_{n, l}(\alpha
   r)^{|l|}e^{-\frac{(\alpha r)^2}{2}}L^{|l|}_n((\alpha
   r)^2)e^{il\theta}\chi_{\sigma}\ ,
\end{equation}
and 
\begin{equation}
E_{n, l,\sigma}=\hbar\Omega(2n + |l| + 1)
 - \hbar\omega_B l + \sigma E_B\ .
\end{equation}
In these equations $n=0, 1, 2,\cdots$ and $l=0,\pm 1,\pm 2, \cdots$ 
are quantum numbers.  $\Omega=\sqrt{\omega^2_0+\omega^2_B}$ 
and $\omega_B=eB/(2m^{\ast})$.
$N_{n,l} = \left(\frac{\alpha^2n!}
{\pi(n+|l|)!}\right)^{\frac{1}{2}}$ with $\alpha
 = \sqrt{m^{\ast}\Omega/\hbar}$. 
$E_B=\frac{1}{2}g\mu_BB$ is the Zeeman splitting
 energy.  $\sigma=\pm1$ refers to the spin polarization
along the $z$-axis. $\chi_\sigma$ represents the eigenfunction of $\sigma_z$.
$L_n^{|l|}$ is the generalized Laguerre polynomial. By solving
\begin{equation}
\label{hp}
H_e|\Psi_\ell\rangle=\epsilon_\ell|\Psi_\ell\rangle\ ,
\end{equation}
 one can determine
the eigenenergy $\epsilon_\ell$ and the eigenfunction of the total electron
system $H_e$. It is noted that due to the presence of the spin-orbit
coupling $H_{so}$, $\sigma$ is no longer a good quantum number. Mixing
occurs for opposite spins:
\begin{widetext}
\begin{equation}
\label{a1}
\langle  n,l,\sigma|H_{so}|n^\prime,l^\prime,\sigma^\prime\rangle
=i2\pi \gamma_c\alpha\delta_{l^\prime+\sigma,l}\delta_{\sigma,-\sigma^\prime}
[\sigma(\omega_B/\Omega) A^{(1)}_{n,n^{\prime},l,l^{\prime}}
-\sigma l^\prime A^{(2)}_{n,n^{\prime},l,l^{\prime}}+A^{(3)}_{n,n^{\prime},l,l^{\prime}}]\ .
\end{equation}
It is this mixing that makes the originally spin-conserving electron-phonon
scattering Eq.\ (\ref{ep}) cause spin relaxation.
\end{widetext} 
$A^{(1)}$ to $A^{(3)}$ in Eq.\ (\ref{a1}) are given in detail in Appendix A.

The eigenfunction $|\Psi_\ell\rangle$ obtained from Eq.\ (\ref{hp}) contains 
spin mixing for each state $\ell$. We assign an eigen state $\ell$ to be
spin-up if $\bar{\sigma}_z=\langle\Psi_\ell|\sigma_z|\Psi_\ell\rangle >0$ 
or spin-down if $\bar{\sigma}_z<0$. 
An electron at initial electron state $i$ with energy $\epsilon_i$ and a
spin polarization can be scattered by the phonon into another
state $f$ with energy $\epsilon_f$ and the {\em opposite} spin
polarization. The rate of such scattering can be 
described by the Fermi golden rule:
 \begin{eqnarray}
   \label{eq:fermirule}
 \Gamma_{i\to f}&=&\frac{2\pi}{\hbar}\sum_{{\bf q}\lambda}|M_{{\bf q}\lambda}|^2|
\langle f|e^{i{\bf q}\cdot{\bf r}}|i\rangle|^2[{\bar n}_{{\bf
   q}\lambda}\delta(\epsilon_f-\epsilon_i-\omega_{{\bf q}\lambda})\nonumber\\
 &&+
 ({\bar n}_{{\bf
   q}\lambda} + 1)\delta(\epsilon_f-\epsilon_i+\omega_{{\bf q}\lambda})]\ ,
 \end{eqnarray}
with ${\bar n}_{{\bf q}\lambda}$ representing the Bose distribution of phonon
with mode $\lambda$ and momentum $q$ at the temperature $T$. Its expression 
after the integration is given in Appendix B.
The SRT $\tau$ can therefore be 
determined by
\begin{equation}
   \label{eq:SRT}
\frac{1}{\tau} = \sum_{i}f_i\sum_f\Gamma_{i\to f}\ ,
 \end{equation}
in which $f_i = C\exp[-\epsilon_i/(k_BT)]$ denotes the Maxwell distribution
of the $i$-th level with $C$ being a constant.

\section{Numerical Results}
We perform a numerical investigation of  the SRT [Eq.\ (\ref{eq:SRT})]
 in  GaAs quantum dots  at low temperatures 
by diagonalizing the Hamiltonian $H_e$ for each given  
dot size: the quantum well width $a$ and the effective diameter 
$d=\sqrt{\frac{\hbar\pi}{m^{\ast}\omega_0}}$, 
as well as each given applied magnetic field $B$. To do so,
we gradually increase the number of basis function
arranged in the order of energy in the 
Hilbert space $|n, l, \sigma\rangle$ to ensure the 0.1\ \% precision
of the converged energy $\epsilon_{\ell}$. As an example for a QD with $B=1$\ T and $a=5$\ nm,
when $d=20$\ nm, in order to converge the lowest 2 (100) levels, one has to
use 12 (120) basis functions;  nevertheless, 
when $d=60$\ nm, in order to converge
the lowest 2 (100) levels, one has to use 20 (200)
basis functions.
\begin{table}[htbp]
\begin{tabular}{lllllll}\hline\hline
D&\mbox{}&$5.3\times10^3$kg/m$^3$&\mbox{}&$\kappa$&\mbox{}&12.9\\
$v_{st}$&\mbox{}&$2.48\times10^3$\ m/s&\mbox{}&g&\mbox{}&$-0.44$\\
$v_{sl}$&\mbox{}&$5.29\times10^3$\ m/s&\mbox{}&$\varXi$&\mbox{}&7.0\ eV\\
$e_{14}$&\mbox{}&$1.41\times10^3$\ V/m&\mbox{}&$m^{\ast}$&\mbox{}&$0.067m_0$\mbox{}\\\hline\hline
\end{tabular}
\caption{Parameters used in the calculation}
\label{tab:parameter}
\end{table}

The electron-phonon scattering is composed of the following
contributions: (i) The electron-acoustic phonon scattering due to the
deformation potential with $M^2_{{\bf q}sl} =
\frac{\varXi^2q}{2DV_{sl}}$; (ii) The electron-acoustic phonon 
scattering due to the piezoelectric field for the longitudinal phonon
mode with $M^2_{{\bf q}pl} =
 \frac{32\pi^2e^2e^2_{14}}{\kappa^2Dv_{sl}}
 \frac{(3q_xq_yq_z)^2}{q^7}$ and  for the two
transverse phonon modes  with  $\sum_{j=1,2}M^2_{{\bf q}pt_j} =
 \frac{32\pi^2e^2e^2_{14}}{\kappa^2Dv_{st}q^5}
[q_x^2q_y^2+q_y^2q_z^2+q_z^2q_x^2-\frac{(3q_xq_yq_z)^2}{q^2}]$.
Here  $\varXi$ stands for the acoustic deformation potential,  $D$ 
is the GaAs volume density, $e_{14}$ represents the
 piezoelectric constant and $\kappa$ denotes the static dielectric
 constant. The acoustic phonon spectra $\omega_{{\bf q}\lambda}$ are 
given by $\omega_{{\bf q}l}= v_{sl}q$ for the longitudinal mode 
and $\omega_{{\bf q}pt} = v_{st}q$ for the transverse mode with 
$v_{sl}$ and $v_{st}$ representing the corresponding sound velocities.
It is noted that notwithstanding the fact that we 
include all these acoustic phonons throughout our computation, for
all the cases we have studied in this paper, 
the main contribution comes from the electron-phonon 
scattering due to the piezoelectric field for the transverse mode with the later
being at least one order of magnitude larger than the other phonon modes.
Moreover, the contribution from LO-phonon is negligible in the
temperature regime we are studying. 
The parameters used in our calculation 
are listed in Table \ref {tab:parameter}.\cite{made}

\subsection{Comparison with previous works at $T=4$\ K\label{sec:compare}} 

We first compare our approach with the  perturbation
calculations widely used in the literature\cite{Woods,Alexander,Alexander1}
at low temperature to double check
the validity of our method as well as that of the perturbation
method  where $H_{so}$ is treated as the perturbation.
Following the previous works,\cite{Woods,Alexander,Alexander1} 
we calculate the SRT between the
lowest two Zeeman splitting levels at $d=20$\ nm and $T=4$\ K. 
Unless specified, the width of the quantum well $a$ is fixed to 
be 5\ nm throughout the paper.
To the first order of $H_{so}$, the energy difference of the 
lowest two states with the opposite spins is 
$\Delta E=2E_B$ (the first order correction is zero) and 
the wave functions of these two states are
\begin{eqnarray}
  \label{eq:wave_lowest}
  \Psi_{\uparrow} &=& \langle r|0,0,\uparrow\rangle\ ,\nonumber\\
  \Psi_{\downarrow} &=& \langle r|0,0,\downarrow\rangle -{\cal A}\langle r|0,1,\uparrow\rangle \ ,
\end{eqnarray}
in which ${\cal A}=i\hbar\gamma_c(\frac{\pi}{a})^2
\frac{\alpha(1-eB/(2\hbar\alpha^2))}{E_{0,1,\uparrow}-E_{0,0,\downarrow}}$.
The SRT $\tau$ is therefore given by  
\begin{eqnarray}
  \label{eq:simplify}
\frac{1}{\tau} 
&=&c|{\cal A}|^2{\bar n_q}q^3\int^{\frac{\pi}{2}}_{0}d\theta\sin^5{\theta}(\sin^4{\theta}+8\cos^4{\theta})\nonumber\\
&&\times\exp(-\frac{1}{2}q^2\sin^2{\theta})I^2(q\cos{\theta})\ ,
\end{eqnarray}
with $q=\Delta E/(\hbar v_{st}\alpha)$,
$c=9\pi \alpha e^2 e^2_{14}/(\hbar D v^2_{st}\kappa^2)$ and
$I(q_z)={8\pi^2\sin(aq_z/{2})}/\{aq_z[4\pi^2-(aq_z)^2)]\}$. 
This is exactly the same 
calculation used in the literature.\cite{Woods,Alexander,Alexander1}

In Fig.\ 1 we compare the SRT  calculated from Eq.\ (\ref{eq:simplify}) (curve with 
$\blacksquare$) with our exact diagonalization method (curve with $\bigcirc$)
described in the previous section, but 
with only the lowest four levels of the Hilbert space, {\em ie.},
$|0,0,\sigma\rangle$ and $|0,1,\sigma\rangle$ ($\sigma=\uparrow$ or $\downarrow$), 
taken, corresponding to the same levels used in the perturbation method. It is seen from the figure that
there is at least one order of magnitude difference between the two curves:
$\tau$ obtained from the perturbation method is much larger than the one from the exact 
diagonalization method. Moreover, the trends of the
magnetic field dependence are also different. We point out that these differences arise from the
fact that only the first order of the perturbation is applied. 
It can be fixed if one further includes the second order correction in $\Delta E$, {\em ie.}
$\Delta E=2E_B + |{\cal A}|^2(E_{0,1,\uparrow}-E_{0,0,\downarrow})$. It is noted that the 
second order correction to the energy difference is much larger than $2E_B$, the unperturbed
energy difference. The SRT calculated from the perturbation method, modified with the energy
correction to the second order, is plotted as a function of the magnetic field $B$ in the same
figure (curve with $\times$). One notices that it exactly hits on the curve
from our diagonalization method (curve with $\bigcirc$). 

 \begin{figure}[htb]
  \psfig{figure=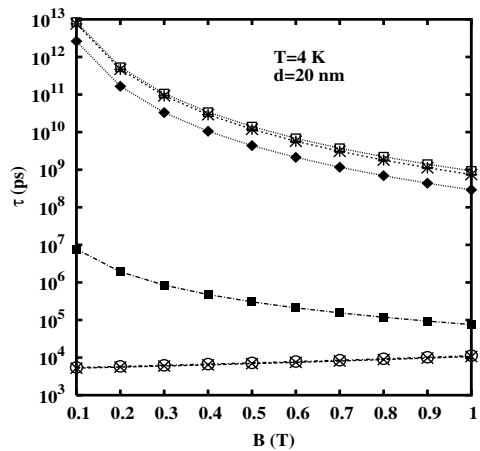,width=7cm,height=6cm,angle=0}
  \caption{SRT versus the magnetic field. Curve with $\blacksquare$:
    perturbation result without energy corrections; Curve with $\bigcirc$:
    diagonalization result but with only the lowest four levels used as
    basis functions; Curve with $\times$: Perturbation result with the
    second  order energy correction;Curve with $\blacklozenge$: exact
    diagonalization result with the energy sufficiently converged;
Curve with $\square$: diagonalization result with the lowest six
levels used as basis functions; Curve with {\LARGE $\ast$}:
perturbation result with the lowest six levels of $H_0$ as basis
functions and with the second order energy correction.}
\label{fig:compare}
\end{figure}

We notice that the diagonalization above includes only the lowest four
levels. If it is adequate in converging the lowest two levels remains
unchecked. As mentioned in the beginning of this section, that for the
size of the dot we are studying here, in order to converge the lowest
two energy levels, one has to use 12 basis functions. $\tau$
calculated from the exact diagonalization method is plotted in also
Fig.\ 1 against the magnetic field (curve with
$\blacklozenge$). Strikingly, it is {\em orders of magnitude} larger
than that from the  perturbation.  

  In order to understand this huge difference, now we include {\em six}
  lowest energy levels of $H_0$,  {\em i.e.}, 
  $|0, 0, \uparrow\rangle$, $|0, 0, \downarrow\rangle$, 
  $|0, 1, \uparrow\rangle$,  $|0, 1, \downarrow\rangle$, 
  $|0, -1, \uparrow\rangle$ and $|0, -1, \downarrow\rangle$,
  as basis functions in the perturbation method. 
  The wavefunctions of the lowest two states of $H_e$ are therefore
  given by
  \begin{eqnarray}
    \label{eq:wave_lowest_6}
    \Psi_{\uparrow}&=&\langle r|0,0,\uparrow\rangle-{\cal B}\langle
    r|0,-1,\downarrow\rangle\nonumber\ ,\\
    \Psi_{\downarrow} &=& \langle r|0,0,\downarrow\rangle -{\cal
      A}\langle r|0,1,\uparrow\rangle \ , 
  \end{eqnarray}
  in which 
  \begin{equation}
    {\cal B}=i\hbar\alpha \gamma_c(\frac{\pi}{a})^2
    \frac{1+eB/(2\hbar\alpha^2)}{E_{0,-1,\uparrow}-E_{0,0,\downarrow}}\ .
  \end{equation}
The energy difference between $\Psi_{\uparrow}$ and
  $\Psi_{\downarrow}$ now becomes
  \begin{equation}
    \Delta E=2E_B+|{\cal
      A}|^2(E_{0,1,\uparrow}-E_{0,0,\downarrow}) - |{\cal
      B}|^2(E_{0,-1,\uparrow}-E_{0,0,\uparrow})\ .
  \end{equation} 
  The corresponding SRT $\tau$ is hence given by
  \begin{eqnarray}
    \label{eq:simplify_6}
    \frac{1}{\tau}
    &=&c|{\cal A}-{\cal B}|^2{\bar n_q}q^3\int^{\frac{\pi}{2}}_{0}d\theta\sin^5{\theta}(\sin^4{\theta}+8\cos^4{\theta})\nonumber\\
    &&\times\exp(-\frac{1}{2}q^2\sin^2{\theta})I^2(q\cos{\theta})\ ,
  \end{eqnarray}
  The numerical results of Eq.\ (\ref{eq:simplify_6}) are plotted by
  the curve with {\LARGE $\ast$} in Fig.\ 1. It is seen from the figure 
  that the inclusion of the additional basis functions in the perturbation 
  method also greatly enhances the SRT by orders of magnitude. The
  results obtained from the exact diagonalization method with the same 
  lowest six levels as basis are given by the curve with $\square$.
  The two curves are almost the same, and much closer to the
  final converged results (curve marked with $\blacklozenge$).

\begin{figure}[htb]
  \psfig{figure=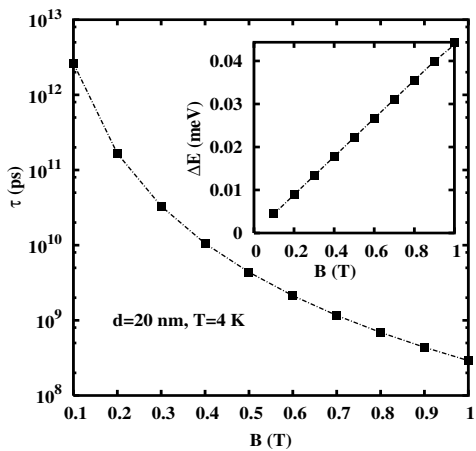,width=7cm,height=6cm}
  \caption{SRT versus the applied magnetic field at d=20~nm and T=4~K. 
    The inset is the corresponding energy splitting $\Delta E$ between the
    lowest two levels}
  \label{fig:energy}
\end{figure}

It is clearly seen from the above calculation that the perturbation
approach widely used in the literature is {\em inadequate} 
in describing the SRT even with the
second order energy corrections included.
 In principal in order to use the perturbation method to
  calculate the SRT, one has to  include sufficient number of the
  states in the basis in stead of only the lowest four levels widely
  used in the literature. This is of course inapplicable especially
  for larger QD's or  higher temperature where one has to include a
  lot of basis functions (for a QD of $d=60$\ nm, one has to use 100
  levels as  basis functions) and the SRT is determined by many levels
(in stead of  only the lowest two) of the total  electron Hamiltonian $H_e$.
Even for the lowest two
levels (for a QD with $d=20$\ nm at 4\ K, the SRT is determined by the
lowest two levels of the total electron Hamiltonian $H_e$), one has to
use a lot of basis functions to converge the energy and the resulting
SRT between these two levels is therefore strongly readjusted. This is
because the spin-orbit coupling is very strong in mixing different
energy levels of $H_0$.

In the following subsections, we therefore reinvestigate the properties
of the SRT based on the exact diagonalization calculation. 

\subsection{SRT of a $d=20$\ nm QD at $T=4$\ K}

As pointed out in the previous subsection that for a QD with $d=20$\ nm,
at $T=4$\ K the SRT is determined by the spin-flip transition between the 
lowest two energy levels after the {\em exact diagonalization},
although at least 12 energy levels in the Hilbert space of $H_0$
are essential in getting these two levels. In this subsection we 
focus on the effects of the external fields on the SRT determined
by these two levels.

It is seen from Eqs.\ (\ref {eq:simplify}) and
  (\ref{eq:simplify_6}) that in a given basis the 
spin relaxation rate $1/\tau$ is determined
by two competing  trends as a function of the energy splitting $\Delta
E$: (i) $q^3{\bar n}_q$, which increases with $\Delta E$ in the present
case, and (ii) $\exp(-\frac{1}{2}q^2\sin^2{\theta})I^2(q\cos{\theta})$,
which decreases with $\Delta E$. Therefore, the SRT can be
uniquely determined by the energy $\Delta E$. For small $\Delta
E$, it is easy to see that the trend (i) dominates the
when $\Delta E\lesssim 7.0\hbar
    v_{st}/d$, which is 0.57\ meV at $d=20$\ nm. That is, the SRT 
decreases with $\Delta E$ when $\Delta E \lesssim 0.57$\ meV.

In Fig.\ {\ref{fig:energy}} the SRT is plotted against the
applied magnetic field $B$.
It is seen from the figure that $\tau$ decreases with the
applied magnetic field. This is understood from the fact that the
energy splitting $\Delta E$ increases with the applied magnetic field
as shown in the inset. Moreover, even for the largest energy splitting
0.04\ meV at $B=1$\ T, it is one order of magnitude smaller than 
0.57\ meV, energy splitting required to have the opposite $\tau$-$B$
dependence.

\begin{figure}[htb]
  \psfig{figure=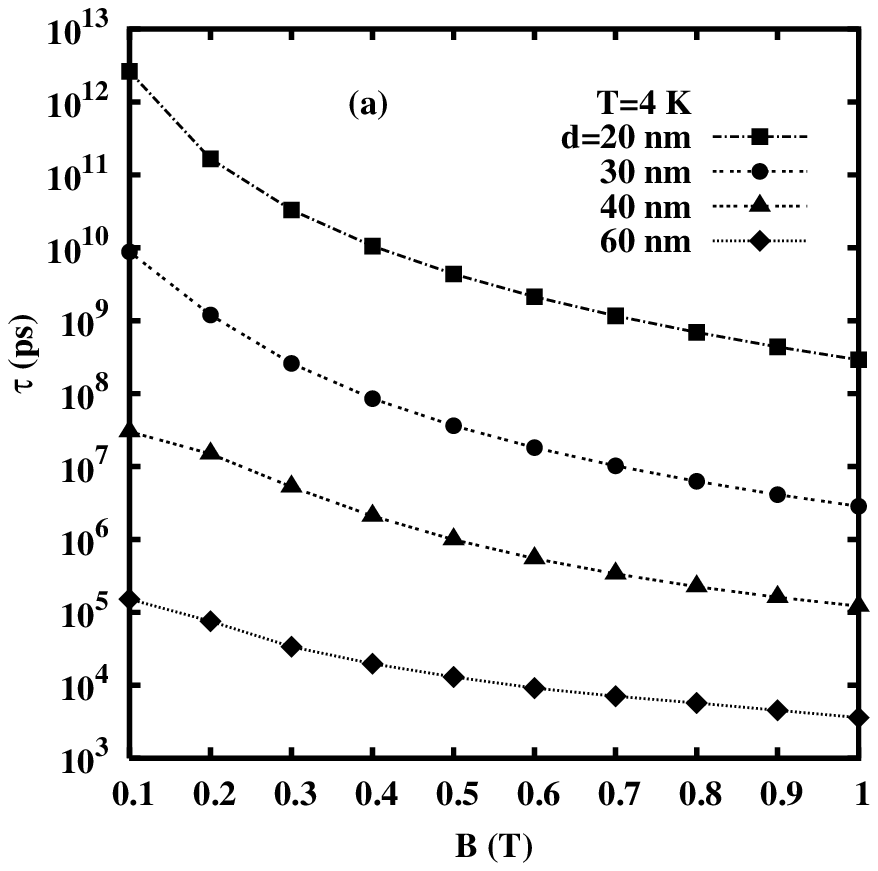,width=7cm,height=6cm,angle=0}
  \psfig{figure=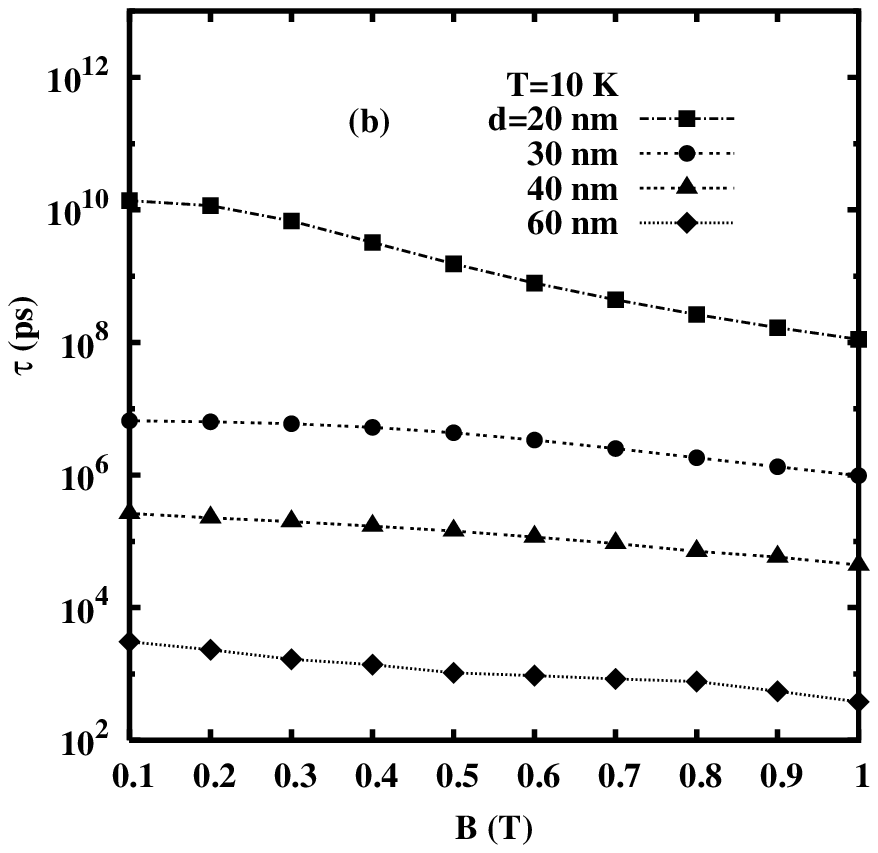,width=7cm,height=6cm,angle=0}
  \caption{The SFT vs. the magnetic field with different sizes of the
  quantum dots:($\blacksquare$) d=~20~nm,($\bullet$) d=~30~nm,
  ($\blacktriangle$) d=~40~nm, and ($\blacklozenge$) d=~60~nm
 for  $T=4$ K (a) and 10\ K (b).} 
  \label{fig:BT}
\end{figure}

\subsection{Magnetic field dependence of the SRT}

We investigate the magnetic field dependence of the SRT for 
different diameters of the QD's at two different temperatures
as shown in Fig.\ \ref{fig:BT}. Unlike the previous subsection
where only the lowest two energy levels are important, here for most
cases one has to include many levels of the total electron Hamiltonian. 

It is seen that the SRT decreases rapidly with
the magnetic field at each dot size and temperature.
This feature is quite opposite to the bulk,\cite{wu} the two-\cite{weng}
 and the one-dimensional\cite{cheng} cases
where the SRT always increases with the magnetic field. This is because
in the dot case there are only discrete energy levels and
the magnetic field helps to increase the spin-flip scattering as
discussed in the previous subsection. Moreover, one notices that the
SRT drops dramatically with the dot size. For a dot with $d=60$\ nm,
the SRT is more than 6 orders of magnitude faster than the one
with $d=20$\ nm. This is understood that for larger dots, more energy levels
are engaged in the spin-flip scattering and hence sharply reduce the SRT.

\begin{figure}[htbp]
  \centering
  \psfig{figure=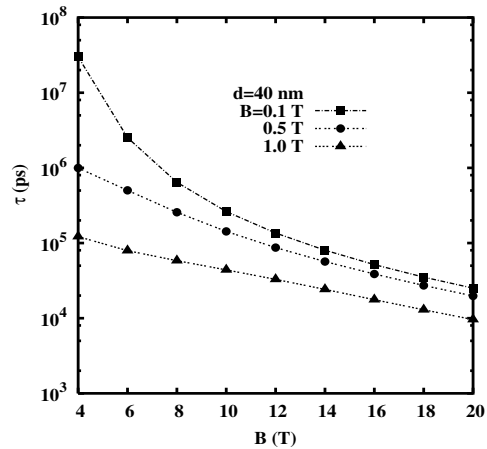,width=7cm,height=6cm,angle=0}
  \caption{The SRT {\em vs}. the temperature 
under different magnetic fields at
 $d=40$\ nm. Curve with $\blacksquare$: $B=0.1$\ T; Curve
with $\bullet$: $B=0.5$\ T; and curve with $\blacktriangle$: $B=1$\ T.}
  \label{fig:TD}
\end{figure}

\subsection{Temperature dependence of the SRT}

We plot the SRT as a function of the temperature in Fig.\ \ref{fig:TD}
for a QD with $d=40$\ nm under three different magnetic fields.
From the figure one finds that the SRT gets smaller with the increase
of the temperature. Moreover, the smaller the magnetic field is, the
faster the SRT drops with the temperature.

These features can be understood as follows:  With the
increase of the temperature, the phonon number ${\bar n}_{{\bf q}\lambda}$
gets larger. This enhances the electron-phonon scattering 
and leads to the larger transition probability.
Moreover, unlike the previous work\cite{Woods} where 
the difference between zero temperature and finite temperatures 
is just the phonon Bose distribution, we stress that
for high temperatures, the occupation to the high energy levels
becomes important and  it is inadequate to consider only the lowest
several levels. For lower magnetic fields, the space between different
energy levels is smaller. Therefore, more levels are included 
in the energy regime determined by $f_i$ in Eq.\ (\ref{eq:SRT})
which leads to a faster response to the temperature. This 
feature is more pronounced in the low temperature regime.
For high temperatures, as there are already many levels included in the
energy space, adding a few more levels does not change the SRT
significantly. Consequently the rates of the decrease of the SRT 
with the temperature become similar for different magnetic fields
when $T>16$\ K.

\subsection{Well width dependence of the SRT}
\begin{figure}[htbp]
  \psfig{figure=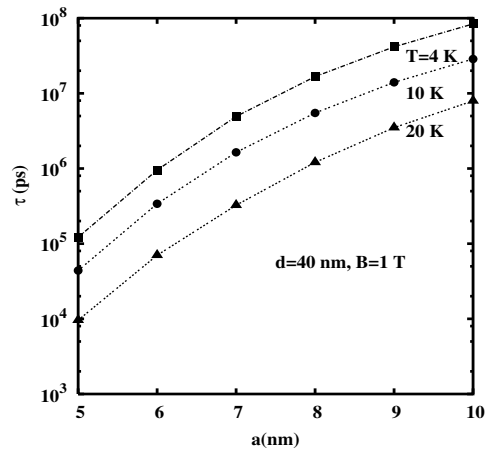,width=7cm,height=6cm,angle=0}
  \caption{The SRT {\em vs.} the width of the quantum well in different
temperatures at B=~1~T. Curve with $\blacksquare$: $T=4$\ K;
Curve with $\bullet$: $T=10$\ K; and Curve with $\blacktriangle$:
$T=20$\ K.}
  \label{fig:AD}
\end{figure}

As the QDs  are confined in the quantum well, it
is necessary to study the quantum well width  dependence of the
SRT as shown in Fig.\ {\ref {fig:AD}} where $\tau$ is plotted
as a function of the well width  $a$ for different
temperatures at B=1\ T. It is noted that the SRT 
increases  with the well width $a$. This is due to the fact that
the spin-orbit coupling $H_{so}$ [Eq.\ (\ref{so})] is 
proportional to $1/a^{2}$. Smaller well width corresponds to
larger spin-orbit coupling and therefore smaller SRT.
We point out here that the well width in the present calculation is 
much smaller than the dot size $d$ and only the lowest subband 
contributes to the SRT. For larger well width, more subbands are
involved and hence there adds an opposite tendency for a shorter SRT
with the increase of the well width.
 
\section{Conclusions}
In conclusion, we have investigated the SRT in GaAs QDs by the exact
diagonalization method with applied magnetic fields. After comparing 
the exact diagonalization method with the
perturbation approach widely used in the literature, we find that the
later is inadequate in accounting for the electron structure and the
SRT in QDs.  This is because that the energy splitting caused by the
spin-orbit coupling is several times large than the Zeeman splitting
used in the perturbation approach. Moreover, a lot more energy levels
of $H_0$ are coupled by the spin-orbit coupling  and therefore
contribute to the lowest energy levels of the total QD Hamiltonian. 
 We therefore reinvestigated the SRT from the exact diagonalization method to
explore its dependence on the magnetic filed, the temperature
and the size of the QD. We find the SRT decreases with the magnetic field, which is 
quite opposite to the bulk, the two- and one- dimensional cases. It also decreases 
with the diameter of the QD, but increases with the width  of the quantum well on which
the QD grows. For high temperature, the SRT becomes much faster due to the stronger
electron-phonon scattering and the wider range of energy space the electron occupies.
 All our investigation suggests the 
importance of the exact calculation of the energy structure.

\acknowledgments
MWW is supported by the  ``100 Person Project'' of Chinese Academy of
Sciences and Natural Science Foundation of China under Grant No.
90303012. He would like to
thank Dr. Marion Florescu for valuable discussion. 
The authors would like to acknowledge fruitful discussions 
with  M. Q. Weng.
\bigskip

\begin{appendix}
\section{The expressions of $A^{(1)}$, $A^{(2)}$ and $A^{(3)}$}

$A^{(1)}$, $A^{(2)}$ and $A^{(3)}$ in Eq.\ (\ref{a1})
are given by
\begin{eqnarray}
A^{(1)}_{n,l,n^{\prime},l^{\prime}}&=&
\alpha\int^{\infty}_{0}r^2R_{n,l}(r)R_{n^{\prime},l^{\prime}}(r)dr\ 
,
\label{aa1}
\\
 A^{(2)}_{n,l,n^{\prime},l^{\prime}}&=&\frac{1}{\alpha}\int^{\infty}_{0}R_{n,l}(r)R_{n^{\prime},l^{\prime}}(r)dr\
 ,
\label{aa2}
  \\
 A^{(3)}_{n,l,n^{\prime},l^{\prime}}&=&\frac{1}{\alpha}\int^{\infty}_{0}
r R_{n,l}(r)\frac{\partial}{\partial r}R_{n^{\prime},l^{\prime}}(r)dr\
,
\label{aa3}
\end{eqnarray}
where 
$R_{n,l} = \sqrt{\frac{\alpha^2 n!}{\pi
    (n+|l|)!}}(\alpha
r)^{|l|}\exp{(-\frac{(\alpha r)^2}{2})}L^{|l|}_n(\alpha^2 r^2)$
is the spatial part of the wave function Eq.\ (\ref{eq:eigen}).  From
the integration over the angular part, we get the relation  
$|l-l^{\prime}| = 1$. Substituting this relation into 
Eqs.\ (\ref{aa1})-(\ref{aa3}), after carrying out the integration we have
\begin{equation}
 A^{(1)}_{n,l,n^{\prime},l^{\prime}} =
 \frac{1}{2\pi}(\sqrt{n+|l|+1}\delta_{n,n^{\prime}} -
 \sqrt{n}\delta_{n^{\prime},n-1} )\ ,
\end{equation}
\begin{equation}
A^{(2)}_{n,l,n^{\prime},l^{\prime}} =\begin{cases}
  \frac{1}{2\pi}\sqrt{\frac{n^{\prime}!(n+|l|)!}
{n!(n^{\prime}+|l^{\prime}|)!}},
&\mbox{if }n^{\prime} > n\\0, &\mbox{otherwise}\end{cases}\ .
\end{equation}
\begin{widetext}
It is noted that due to the symmetry between \{n, l\} and
\{$n^{\prime}$, $l^{\prime}$\}, in above two equations we 
only give the results with $|l^{\prime}| = |l| + 1$.
Finally
\begin{eqnarray}
A^{(3)}_{n,l,n^{\prime},l^{\prime}} &&=
  |l^{\prime}|A^{(2)}_{n,l,n^{\prime},l^{\prime}} +
  A^{(1)}_{n,l,n^{\prime},l^{\prime}}
-\begin{cases}\frac{\sqrt{n+|l|}}{\pi},&\mbox{if }|l^{\prime}|=|l|-1
  \mbox{ and
  }n^{\prime}=n\\\frac{|l^{\prime}|}{\pi}\sqrt{\frac{n^{\prime}!(n+|l|)!}{n!(n^{\prime}+|l^{\prime}|)!}},&\mbox{if }|l^{\prime}|=|l|+1\mbox{ and }n^{\prime}\geqslant n\\\frac{\sqrt{n}}{\pi},&\mbox{if }|l^{\prime}|=|l|+1\mbox{ and  }n^{\prime}=n-1\\0, &\mbox{otherwise}\end{cases} \ .
\end{eqnarray}

\section{The expression of $\Gamma_{i\to j}$}
\begin{equation}
\label{eq:last_gamma}
  \Gamma_{i\to f}=\sum_{\lambda}
  \frac{1}{(2\pi)^2v_{\lambda}}N_q\int^{q}_0dQ\frac{qQ}{q_z}\exp{(-\frac{Q^2}{2\alpha^2})}
G^2_{i, f}(\frac{Q^2}{(2\alpha)^2},
  q_z)\int^{2\pi}_0d\theta|M_{q,\lambda}|^2
\end{equation}
with
${\bf q} = (Q\cos\theta, Q\sin\theta, q_z)$ and $q=|{\bf
  q}|=\frac{|E_i-E_j|}{v_{\lambda}}$.
Here $N_q ={\bar n}_q$ if $E_i>E_j$ or ${\bar n}_q +1$ if $E_i<E_j$. 
$G_{i,f}$ in Eq.\ (\ref{eq:last_gamma}) is
\begin{equation}
  \label{eq:element}
G_{i, f}(Q^2/(4a^2), q_z)= \sum_{n_1,l_1,n_2,l_2,\sigma} C_{n_1,l_1,\sigma}^{i}
(C_{n_2,l_2,\sigma}^{f})^\ast \langle
{n_2,l_2}|\exp{[i(q_xx+q_yy)]}|{n_1,l_1}\rangle\exp{(\frac{Q^2}{2\alpha^2})}
I(q_z)\ ,
\end{equation}
in which
\begin{eqnarray}
&&{\exp{(\frac{Q^2}{2\alpha^2})}}\langle {n_2,l_2}|\exp{[i(q_xx+q_yy)]}|{n_1,l_1}
\rangle=\\
&&\sqrt{\frac{n_1!n_2!}{(n_1+|l_1|)!(n_2+|l_2|)!}}
{e^{i(l_1-l_2)(\frac{\pi}{2}+\theta)}}{(\mbox{sgn}(l_1-l_2)\frac{Q}{2\alpha})}^{|l_1-l_2|}\sum^{n_1}_{i=0}
\sum^{n_2}_{j=0}{\cal C}^{i}_{n_1,|l_1|}{\cal C}^{j}_{n_2,
  |l_2|}n!L^{|l_1-l_2|}_n(\frac{Q^2}{(2\alpha)^2})\nonumber
\end{eqnarray}
with sgn$(x)$ denoting the sign function, ${\cal C}_{n,l}^m=\frac{(-1)^m}{m!}
\genfrac{(}{)}{0pt}{}{n+l}{n-m}$ and $n=i+j+\frac{|l_1|+|l_2|-|l_1-l_2|}{2}$.
\end{widetext}
\end{appendix}


\end{document}